\documentclass[english,12pt]{article}
\usepackage{array}
\usepackage{graphicx}
\usepackage{amssymb}
\usepackage[english]{babel}
\usepackage{amsmath}
\usepackage{multirow}
\usepackage{prettyref}
\usepackage{babel}
\usepackage{units}
\usepackage[latin1]{inputenc}
\usepackage{amsfonts}
\usepackage{amssymb}
\usepackage{babel}
\usepackage{color}
\usepackage{cite}
\def\@fmsl@sh#1#2#3{\m@th\ooalign{$\hfil#1\mkern#2/\hfil$\crcr$#1#3$}}
 \def\eq#1\en{\begin{equation}#1\end{equation}}
\def\s[#1,#2]{[#1\stackrel{\star}{,}#2]}
\def\sx[#1,#2]{[#1\stackrel{\star_{x}}{,}#2]}

\def\bc{\begin{center}}
\def\ec{\end{center}}

\def\gsim{\mathrel{\mathpalette\atversim>}}

\def\bc{\begin{center}}
\def\ec{\end{center}}

\def\gsim{\mathrel{\rlap{\lower4pt\hbox{\hskip1pt$\sim$}}

    \raise1pt\hbox{$>$}}}       

\def\gsim{\mathrel{\rlap{\lower4pt\hbox{\hskip1pt$\sim$}}
    \raise1pt\hbox{$>$}}}       

\begin{document}
\makeatletter
\def\fmslash{\@ifnextchar[{\fmsl@sh}{\fmsl@sh[0mu]}}
\def\fmsl@sh[#1]#2{%
  \mathchoice
    {\@fmsl@sh\displaystyle{#1}{#2}}%
    {\@fmsl@sh\textstyle{#1}{#2}}%
    {\@fmsl@sh\scriptstyle{#1}{#2}}%
    {\@fmsl@sh\scriptscriptstyle{#1}{#2}}}
\def\@fmsl@sh#1#2#3{\m@th\ooalign{$\hfil#1\mkern#2/\hfil$\crcr$#1#3$}}
\makeatother

\thispagestyle{empty}
\begin{titlepage}
\boldmath
\begin{center}
  \Large {\bf Quantum Hair in Electrodynamics and Gravity}
    \end{center}
\unboldmath
\vspace{0.2cm}
\begin{center}
{{\large Xavier~Calmet}\footnote{E-mail: x.calmet@sussex.ac.uk}$^a$
{\large and}  {\large  Stephen D. H. Hsu}\footnote{E-mail: hsusteve@gmail.com}$^b$}
 \end{center}
\begin{center}
$^a${\sl Department of Physics and Astronomy, \\
University of Sussex, Brighton, BN1 9QH, United Kingdom
}\\
$^b${\sl Department of Physics and Astronomy\\ Michigan State University, East Lansing, Michigan 48823, USA}\\

\end{center}
\vspace{2cm}
\begin{abstract}
\noindent
We demonstrate the existence of quantum hair in electrodynamics and gravity using effective action techniques. In the case of electrodynamics we use the Euler-Heisenberg effective action while in the case of quantum gravity we use the unique effective action.  We give a general formulation of these effects which applies to both theories and discuss analogies and differences between them. Furthermore, we present a QED analog to black hole evaporation. Spontaneous pair production in the external field of a ball of charge is analogous to Hawking radiation from black holes. Assuming spherical symmetry, the Gauss law prevents the external field from depending on the density profile of the ball. Quantum corrections violate these expectations, showing that quantum radiation can encode classically forbidden information about the source.

\end{abstract}
\vspace{5cm}
\end{titlepage}

\section{Introduction}

In this paper we demonstrate that quantum hair exists in quantum gravity for any source energy momentum tensor. By quantum hair we mean quantum corrections to classical solutions in general relativity describing the exterior metric. These quantum corrections can carry information about the interior quantum state which would otherwise be forbidden by the no-hair theorem, and hence play an important role in the black hole information paradox \cite{Calmet:2021stu,Calmet:2021cip,Calmet:2022swf}. 

Our argument is independent of the ultra-violet theory of quantum gravity and only relies on well established methodology using the unique effective action \cite{Barvinsky:1983vpp,Barvinsky:1985an,Barvinsky:1987uw,Barvinsky:1990up,Buchbinder:1992rb}. We show that this kind of hair exists as well in quantum electrodynamics (QED) using the well-known Euler-Heisenberg effective action. The presence of hair in both QED and quantum gravity is linked to a violation of the Gauss law by the field configurations which solve the (quantum corrected) field equations deduced from in the effective field theory.  Furthermore, we show that in both theories, there is a quantum memory effect. The external gravitational or photon field stores information from the dynamical history of the matter or charge source.

Note, the violation of Gauss law we describe here is in the form of solutions of the field equations derived from the quantum effective action. This is not to be confused with the Gauss law constraint imposed in the quantization of QED and quantum gravity. In classical field theory in 3+1 dimensions, the Gauss law requires that the potential of a point source have $1/r$ dependence and the corresponding force field, which is the gradient of this potential, behave as $1/r^2$. Quantum effects in both theories lead to deviations from these behaviors. Consequently the external field of, e.g., a spherically symmetric source is not as constrained as in the original classical theory (cf Birkhoff's theorem in relativity) \cite{Calmet:2017qqa}.

\section{Gauss law in QED}
Much of this section is a review of the familiar results presented in \cite{BermudezManjarres:2020hpk}. Reviewing this work is useful to highlight the connections between QED and quantum gravity.

We consider QED at very low energies: we are interested in photons that have energies much smaller than the electron mass. This theory has a long history and goes back to Euler and Heisenberg who first derived the effective action. The effective action can be obtained by integrating out the electron from the path integral, which leads to non-linearity in the theory. One obtains (see e.g. \cite{BermudezManjarres:2020hpk}) the following action
\begin{eqnarray}\label{SQED}
S&=&\int d^4x\left( -\frac{1}{4} F_{\mu\nu}F^{\mu\nu}+\frac{e^4}{360 \pi^2 m_e^4}\left[ \left(F_{\mu\nu}F^{\mu\nu}\right)^2+\frac{7}{4}\left(F_{\mu\nu}\tilde F^{\mu\nu}\right)^2\right] \right .
 \nonumber \\ && \left .+\frac{e^2}{360 \pi m_e^2}\left[\left(\partial_\alpha F^{\alpha}_{\ \beta} 
\partial_\nu F^{\nu\beta}\right)+ F_{\mu\nu} \Box F^{\mu\nu}\right]+ e A_\mu j^\mu  \right ), 
\end{eqnarray}
where the term quadratic in the field strength is the famous Euler-Heisenberg term. It is proportional to $e^4$ and suppressed by four powers of the electron mass. The quantum correction which is second order in the field strength (and proportional to $e^2$) comes from the photon polarization diagram. Parts of this operator come from expanding the dimension four operator $F_{\mu\nu}\log((m_e^2+\Box)/m_e^2)F^{\mu\nu}$ for energies small in comparison to the electron mass. 

We can write a formal solution to QED field equations
 \begin{eqnarray}
 A_\mu(x)=\int \frac{d^4k}{(2\pi)^4} \int d^4x^\prime D_{\mu\nu}(k) 
 e^{i k(x-x^\prime)} J^\nu(x^\prime)
 \end{eqnarray}
with
 \begin{eqnarray}
 D_{\mu\nu}(k) =\frac{1}{k^2-\Pi(k^2)} \eta_{\mu\nu}.
 \end{eqnarray}
 In classical electrodynamics $\Pi(k^2)=0$ and the Gauss law applies as can be seen trivially:
  \begin{eqnarray}
 E_i(x)=\nabla_i A_0(x)&=&\nabla_i \int \frac{d^4k}{(2\pi)^4} \int d^4x^\prime \frac{1}{k^2} e^{i k(x-x^\prime)} J_0(x^\prime)
 \nonumber \\ &=&\frac{1}{4 \pi \epsilon_0} \int \rho(x_i^\prime)
\frac{x_i-x^\prime_i}{|\Vec{x}-\Vec{x^\prime}|^3} d^3 x_i^\prime ~~.
 \end{eqnarray}
 For a spherically symmetric charge distribution one obtains
 \begin{eqnarray}
 E_i(x)= \frac{Q}{4 \pi \epsilon_0} \frac{\hat x_i}{|\Vec{x}|^2}
 \end{eqnarray}
so that the electric field only depends on the total charge in the region defined by the charge density. However, when quantum electrodynamics corrections are taken into account, $\Pi(k^2)$ is a function which has a non-trivial $k^2$ dependence and the Fourier transform of the charge density is modified. Note that $\Pi(k^2)$ can be obtained from the field equations for the photon derived from Eq. (\ref{SQED}). The Euler-Heisenberg part of the action thus contributes as well to $\Pi(k^2)$.

We can expand  $D_{\mu\nu}(k)$ in $1/k^2$ and it is clear that two sources $J^\mu$ with the same overall $Q$ but different space-time distributions will lead to different field configurations $A_\mu$ \cite{FN1}. This is analogous to the quantum hair identified in the case of gravity which we discuss in the next section. Note that in the case of quantum gravity, screening is not possible because there are no negative masses, while it may be possible in QED as there are both negative and positive charges in the theory. This is a notable difference between the two theories.

We can make the violation of the Gauss law explicit by looking at the electric field of a point charge for the effective action for QED above is given by \cite{BermudezManjarres:2020hpk}
 \begin{eqnarray}
 E(r)= 
 \frac{e}{r^2}-\frac{2 e^7}{45 \pi m_e^4 r^6} +\frac{56 e^9}{2025 \pi m_e^6 r^8} 
 -\frac{112 e^{11}}{ 3375 \pi m_e^8 r^{10}}
 + \frac{4 e^{13}}{675 \pi^2 m_e^8 r^{10}}  + \dots 
\end{eqnarray}
The leading QED correction comes from the Euler-Heisenberg term, and all contain powers of Planck's constant $\hbar$. The electric field can be derived from the potential
\begin{eqnarray}
 \phi(r)= 
 \frac{e}{r}-\frac{2 e^7}{225 \pi m_e^4 r^5} +\frac{8 e^9}{2025 \pi m_e^6 r^7} 
 -\frac{112 e^{11}}{ 30375 \pi m_e^8 r^{9}}
 + \frac{4 e^{13}}{6075 \pi^2 m_e^8 r^{9}} +  \dots .
\end{eqnarray}
The Gauss law only holds for the classical contribution (i.e., the first term in the expansion) but not in the full quantum theory. For a spherically symmetric source the quantum electrodynamics corrections become dependent on the charge density profile.

\section{Quantum hair in gravity for a generic matter distribution}
In this section, we show that quantum hair exists for any matter distribution, that is for any energy-momentum tensor. We then discuss the connection between quantum hair and the Gauss law in gravity and highlight the similarities to QED.

We work with the Vilkovisky-DeWitt unique effective action of quantum gravity at second order in curvature \cite{Barvinsky:1983vpp,Barvinsky:1985an,Barvinsky:1987uw,Barvinsky:1990up,Buchbinder:1992rb}. It is given by
\begin{equation}
	\Gamma_{\rm QG} = \Gamma_{\rm L} + \Gamma_{\rm NL} \, ,
\end{equation}
where the local part of the effective action is given by
\begin{align}
	\Gamma_{\rm L} 
	&= 
	\int d^4x \, \sqrt{|g|} \left[ \frac{M_P^2}{2}
	\big(\mathcal{R} - 2\Lambda\big)
	+ c_1(\mu) \, \mathcal{R}^2 
	+ c_2(\mu) \, \mathcal{R}_{\mu\nu} \mathcal{R}^{\mu\nu} 
	\right.\nonumber\\
	&\qquad \qquad \qquad
	+ c_{3}(\mu) \, \mathcal{R}_{\mu\nu\rho\sigma} \mathcal{R}^{\mu\nu\rho\sigma} 
	+ c_4(\mu) \, \Box \mathcal{R}
	+ \mathcal{O}(M_P^{-2}) \Big]
\end{align}
with the Planck mass $M_P=\sqrt{\hbar/ G_N}$, and the non-local part of the action is given by
\begin{align}
	\Gamma_{\rm NL} &= 
	- \int d^4x \, \sqrt{|g|} \left[ 
	\alpha \, \mathcal{R} \ln \left(\frac{\Box}{\mu^2} \right) \mathcal{R}
	+ \beta \, \mathcal{R}_{\mu\nu} \ln \left( \frac{\Box}{\mu^2} \right) \mathcal{R}^{\mu\nu}
	\right.\nonumber\\
	&\qquad \qquad \qquad \qquad \left.
	+ \gamma \, \mathcal{R}_{\mu\nu\rho\sigma} \ln \left( \frac{\Box}{\mu^2} \right) \mathcal{R}^{\mu\nu\rho\sigma}
	+ \mathcal{O}(M_P^{-2}) \right].
\end{align}
We will set the cosmological constant $\Lambda=0$ for simplicity, and we ignore the boundary term associated with $c_4$, as it does not contribute to Einstein's equations. Furthermore, using the local and non-local Gauss-Bonnet identities \cite{Calmet:2018elv}, we obtain
\begin{align}
	\Gamma_{\rm QG} 
	&=
	\int d^4x \, \sqrt{|g|} \left[ 
	\frac{M_P^2}{2} \,\mathcal{R}
	+ \tilde{c}_1(\mu) \, \mathcal{R}^2 
	+ \tilde{c}_2(\mu) \, \mathcal{R}_{\mu\nu} \mathcal{R}^{\mu\nu} 
	+ \tilde{\alpha} \, \mathcal{R} \ln \left(\frac{\Box}{\mu^2} \right) \mathcal{R}
	\right. \nonumber\\
	& \qquad \qquad \qquad \left.
	+ \tilde{\beta} \, \mathcal{R}_{\mu\nu} \ln \left( \frac{\Box}{\mu^2} \right) \mathcal{R}^{\mu\nu}
	+ \mathcal{O}(M_P^{-2}) \right]
\end{align}
with $\tilde{c}_1 = c_1 - c_3$, $\tilde{c}_2 = c_2 + 4 c_3$, $\tilde{\alpha} = \alpha - \gamma$ and $\tilde{\beta} = \beta + 4 \gamma$.

The field equations derived from the unique quantum gravitational effective action  at second order in curvature are thus given by 
\begin{eqnarray} \label{FEq}
{\cal R}_{\mu\nu} - \frac{1}{2}\, {\cal R}\, g_{\mu\nu} - 16\,\pi\,G_N \left( H_{\mu\nu}^{\rm L} + H_{\mu\nu}^{\rm NL} \right)
= 8 \, \pi \, G_N T_{\mu\nu}
\ ,
\end{eqnarray}
where $G_N$ is Newton constant, $T_{\mu\nu}$ is the energy-momentum tensor and with
\begin{align}
H_{\mu\nu}^{\rm L} 
=
&\,  
\tilde{c}_1
\left( 2\, {\cal R}\, {\cal R}_{\mu\nu} - \frac{1}{2}\, g_{\mu\nu}\, {\cal R}^2 + 2\, g_{\mu\nu}\, \Box {\cal R} - 2 \nabla_\mu \nabla_\nu {\cal R}\right) 
\label{eq:EQMLoc}
\\
&\,
+\tilde{c}_2
\left( 2\, {\cal R}_{~\mu}^\alpha\, {\cal R}_{\nu\alpha} - \frac{1}{2}\, g_{\mu\nu}\, {\cal R}_{\alpha\beta}\, {\cal R}^{\alpha\beta}
+ \Box {\cal R}_{\mu\nu} + \frac{1}{2}\, g_{\mu\nu}\, \Box {\cal R}
- \nabla_\alpha \nabla_\mu {\cal R}_{~\nu}^\alpha
- \nabla_\alpha \nabla_\nu {\cal R}_{~\mu}^\alpha \right)
\ ,
\nonumber
\end{align}
and
\begin{align}
H_{\mu\nu}^{\rm NL} 
=
&\,
 - 2\,\tilde{\alpha}
 \left( {\cal R}_{\mu\nu} - \frac{1}{4}\, g_{\mu\nu}\, {\cal R}
 + g_{\mu\nu}\, \Box
 - \nabla_\mu \nabla_\nu \right)
 \ln\left(\frac{\Box}{\mu^2}\right)\, {\cal R}
 \nonumber
 \\
&\, 
- \tilde{\beta}
\bigg( 2\, \delta_{(\mu}^\alpha\, {\cal R}_{\nu)\beta}
- \frac{1}{2}\, g_{\mu\nu}\, {\cal R}_{~\beta}^\alpha
+ \delta_{\mu}^\alpha\, g_{\nu\beta}\, \Box
+ g_{\mu\nu}\, \nabla^\alpha \nabla_\beta    \\ \nonumber 
&\quad 
- \delta_\mu^\alpha\, \nabla_\beta \nabla_\nu
- \delta_\nu^\alpha\, \nabla_\beta \nabla_\mu 
\bigg)
\ln\left(\frac{\Box}{\mu^2}\right)\, {\cal R}_{~\alpha}^\beta. 
\end{align}
Eqs. (\ref{FEq}) can be solved in perturbation theory: let $\tilde{g}_{\mu\nu} = g_{\mu\nu} + g_{\mu\nu}^{\rm q}$ where $g_{\mu\nu} $ \cite{FN} is the classical solution, and compute $g_{\mu\nu}^{\rm q}$, the quantum correction. In these equations the $\log \Box R^{\mu...\nu}_{\alpha... \beta}$ terms are represented as kernels that are integrated over curvature terms which are functions of the energy-momentum tensor (see e.g. \cite{Calmet:2021stu,Calmet:2019eof,Calmet:2017qqa,Calmet:2018rkj,Codello:2015pga,Donoghue:2014yha,Satz:2004hf}):
\begin{eqnarray}
\log \Box R^{\mu...\nu}_{\alpha... \beta}=\int_{-\infty}^{\infty}d^4 y \langle x|\log \Box|y\rangle R^{\mu...\nu}_{\alpha... \beta}(y),
\label{RT}
\end{eqnarray}
where the integral is over space and time. Using the field equations and perturbation theory, one can express $R^{\mu...\nu}_{\alpha... \beta}(y)$ in terms of the energy-momentum tensor. The boundaries of the integral are fixed by the extension in space and time of the energy-momentum tensor. However, the kernel truncates the integral for values of $y_0$ larger than $x_0$, so that one only integrates over past events and not towards the future, see, e.g., \cite{Calmet:2018rkj}.

In general one will have an interior $T^{\mu\nu}$ metric solution and an exterior one. Eq. (\ref{RT}) and the fact that we can re-express $R^{\mu...\nu}_{\alpha... \beta}(y)$ in terms of the energy-momentum tensor implies that the quantum corrections to the outside metric due to $H_{\mu\nu}^{\rm NL}$ terms must depend on $T^{\mu\nu}$ whatever the matter distribution might be, i.e. a static star, a collapsing star or a black hole. The outside metric will always retain a memory of the interior of the matter distribution.  Furthermore, any deviation from the $1/r$ Newtonian potential will lead to quantum hair whether these corrections are generated by the local or non-local part of the action. Thus, the quantum hair must be present for any gravitational bound state and for black holes in particular. This implies that the Gauss law (as applied to quantum corrected fields in the effective action) is also violated in quantum gravity as we shall discuss in the next section.

Our result implies in particular that the no-hair theorem does not apply to black holes \cite{Calmet:2021stu}. Note that in classical gravity, the no-hair theorem applies only after the black hole has settled in a stable classical configuration, i.e., after a transient period during which it radiates away information beyond $M$, $J$ and $Q$. What happens to static vacuum solutions in quantum gravity? In quantum gravity, black holes are strictly speaking never vacuum solutions: the energy momentum tensor cannot be zero as space-time is filled by the radiation emitted by the black hole as it stabilises. There will thus always be some quantum gravitational corrections to the $1/r$ potential. Even if we considered the unphysical situation where the exterior energy-momentum tensor is exactly zero, it is known that corrections to the Schwarzschild metric appear at third order in curvature \cite{Calmet:2021lny} due to the local part of the effective action. Thus the no-hair theorem does not apply in quantum gravity.

\section{Gauss law in Quantum Gravity}
We can now highlight another interesting connection between the Gauss law in QED and in quantum gravity. We wrote a formal solution to field equations of low energy QED, we can also explicitly write a formal expression for the field equations resulting from the effective action for quantum gravity:
\begin{eqnarray}
 h_{\mu\nu}(x)=\int \frac{d^4k}{(2\pi)^4} \int d^4x^\prime D_{\mu\nu\alpha\beta}(k) 
 e^{i k(x-x^\prime)} T^{\alpha\beta}(x^\prime)
 \end{eqnarray}
with
  \begin{eqnarray}
 D_{\mu\nu\alpha\beta}(k) =\frac{1}{k^2-\Pi(k^2)} (L_{\alpha\mu}L_{\beta\nu}+L_{\alpha\nu}L_{\beta\mu}-L_{\mu\nu}L_{\alpha\beta})
 \end{eqnarray}
and
  \begin{eqnarray}
  L_{\mu\nu}= \eta_{\mu\nu}-\frac{p_\mu p_\nu}{p^2}.
\end{eqnarray}
Again, $\Pi(k^2)$ can be obtained from the effective action.
At tree level $\Pi(k^2)=0$, resulting in a $1/r$ potential and Gauss law. However at the quantum level $\Pi(k^2)$ and the source $T^{\mu\nu}$ receive corrections. The gravitational potential receives quantum corrections which in general lead to terms in the potential $\sim r^{-n}$. In the case of a dust ball the leading correction is of the form $r^{-5}$ \cite{Calmet:2021stu,Calmet:2019eof}.

In other words, there is quantum hair for any $T^{\mu\nu}$ and the quantum hair in gravity is linked to the violation of the Gauss law by the quantum corrected exterior fields. The hair in QED and quantum gravity have the same origin, namely quantum corrections which modify the field equations.

\section{Quantum memory in QED}
Let us point out yet another interesting analogy between QED and quantum gravity. 
The term $F_{\mu\nu} \Box F^{\mu\nu}$ in the effective action has a mathematical structure that resembles very much that of the operators of the type  $R \log{\Box} R$ which gives rise to the quantum memory effect in the case of quantum gravity discussed in section 2. The quantum memory effect identified in quantum gravity is also present in QED. The derivative couplings in the QED effective action at ${\cal O}(e^2)$ also require integration over some space-time dependent bilocal kernels. Indeed the action of the d'Alembertian on the field strength tensor must be understood as a non-local term, because the kernel is non-local: $\langle x|\Box|y\rangle=L(x-y)=\int d^4q/(2\pi)^4 e^{-i q\cdot(x-y)}(-q^2)$. When solving the field equations resulting from the effective action using perturbation theory around a classical solution, one will have to integrate over the whole space-time history of the physical configuration (but again not towards the future), leading to a memory effect as in the case of quantum gravity.  Indeed, as in the case of quantum gravity, in perturbation theory, $F_{\mu\nu}$ can be related to $J_\mu$ using the field equations. Thus when solving the quantum corrected field equations, one ends up with terms of the type $\int dy^0 dy^i \langle x|\Box|y\rangle J_\mu(y^0,y^i)$ where the $y^0$ integral is cut off at $x^0$.
Non-local effects in low energy QED have been discussed previously in \cite{Donoghue:2015nba,Donoghue:2015xla}. Note though, that in these papers, the authors analyzed different operators because they had considered a model with massless fermions which were integrated out of the theory. In contrast, we have considered massive electrons here and thus a realistic theory. However, the overall conclusions are the same.
   
\section{QED black hole analog and pseudo-Hawking radiation}

In classical electromagnetism the external electric field of a spherically symmetric charge distribution only depends on its total charge $Q$. As mentioned above this is a specific consequence of the $1/r$ form of the electrostatic potential, or equivalently the $1/r^2$ fall-off of the electric field.

The Schwinger mechanism in a background electric field describes the pair production of electrons and positrons from the vacuum. Via this process, the spherically symmetric charge distribution with charge $+Q$ can ``evaporate'' by emitting positrons and absorbing electrons, gradually decreasing $Q$ to zero. This is an analog of Hawking radiation:

1. The electrons and positrons are produced in a paired quantum state with the same quantum numbers as the vacuum, analogous to Hawking radiation.

2. At the classical level, the external electric field contains no information about the density profile $\rho(r)$ of the charge distribution. The classical no-hair theorem similarly constrains the black hole external metric. 

However, due to the entirely quantum mechanical effect which produces the Euler-Heisenberg correction to the potential $\delta \phi (r) \sim 1/r^5$, the external field in which pair production occurs does depend on $\rho (r)$. Thus, the evaporation amplitudes {\it can} depend on the radial charge distribution $\rho (r)$, not just the total charge $Q$. Classical reasoning leads to incorrect conclusions about information loss in the radiation process, just as in the Hawking paradox \cite{Calmet:2021cip}. 

\section{Conclusions}
The quantum effective actions for both electrodynamics and gravity lead to field equations which couple a compact source (charge current or energy-momentum tensor) to external fields (electromagnetic or graviton field) in a manner which, generically, leads to quantum memory and quantum hair effects. External solutions of the field equations deviate, due to quantum corrections, from the familiar classical forms that satisfy the Gauss law. As a specific consequence, more information about the interior source configuration is encoded in the external field than in the classical theory. 

As specific applications, we considered semiclassical sources (large black hole, macroscopic charge distribution), which allowed us to solve the quantum corrected field equations by expanding around a classical solution. However, fully quantum statements regarding quantum hair are also possible, which do not, for example, require a semiclassical source. In \cite{Calmet:2021stu,Calmet:2021cip,Calmet:2022swf} it was shown that the quantum state of a compact source (e.g., in an energy eigenstate or superposition thereof) determines certain aspects of the quantum state of its external field. In principle, measurements of the external fields can fully determine the interior state of a black hole.

{\it Acknowledgments}: The work of X.C. is supported in part  by the Science and Technology Facilities Council (grants numbers ST/T00102X/1 and ST/T006048/1).

{\it Data Availability Statement:}
This manuscript has no associated data. Data sharing not applicable to this article as no datasets were generated or analysed during the current study.


\begin{thebibliography}{0}
\bibitem{Calmet:2021stu}
X.~Calmet, R.~Casadio, S.~D.~H.~Hsu and F.~Kuipers,
Phys. Rev. Lett. \textbf{128}, no.11, 111301 (2022)
[arXiv:2110.09386 [hep-th]].

\bibitem{Calmet:2021cip}
X.~Calmet and S.~D.~H.~Hsu,
Phys. Lett. B \textbf{827} (2022), 136995
[arXiv:2112.05171 [hep-th]].

\bibitem{Calmet:2022swf}
X.~Calmet and S.~D.~H.~Hsu,
EPL \textbf{139} (2022) no.4, 49001
[arXiv:2207.08671 [hep-th]].

\bibitem{Barvinsky:1983vpp}
A.~O.~Barvinsky and G.~A.~Vilkovisky,
Phys. Lett. B \textbf{131}, 313-318 (1983).
  
\bibitem{Barvinsky:1985an} 
A.~O.~Barvinsky and G.~A.~Vilkovisky,
Phys.\ Rept.\  {\bf 119}, 1 (1985).
  
\bibitem{Barvinsky:1987uw} 
A.~O.~Barvinsky and G.~A.~Vilkovisky,
Nucl.\ Phys.\ B {\bf 282}, 163 (1987).
  
\bibitem{Barvinsky:1990up} 
A.~O.~Barvinsky and G.~A.~Vilkovisky,
Nucl.\ Phys.\ B {\bf 333}, 471 (1990).

\bibitem{Buchbinder:1992rb} 
I.~L.~Buchbinder, S.~D.~Odintsov and I.~L.~Shapiro,
``Effective action in quantum gravity,''
(CRC Press, Bristol, 1992)




\bibitem{Calmet:2017qqa}
X.~Calmet and B.~K.~El-Menoufi,
Eur. Phys. J. C \textbf{77}, no.4, 243 (2017)
[arXiv:1704.00261 [hep-th]].


\bibitem{BermudezManjarres:2020hpk}
A.~D.~Berm\'udez Manjarres, M.~Nowakowski and D.~Batic,
Int. J. Mod. Phys. A \textbf{35} (2020) no.33, 2050211
[arXiv:2012.02664 [hep-ph]].



\bibitem{FN1} Note that, given the current state of the art, solutions can only be obtained in the semiclassical approximation as they are obtained by solving the quantum corrected field equations in perturbation theory around the classical solution.




\bibitem{Calmet:2018elv}
X.~Calmet,
Phys. Lett. B \textbf{787}, 36-38 (2018)
[arXiv:1810.09719 [hep-th]].

\bibitem{FN} As in the QED case, we make a semiclassical approximation at this stage by solving the quantum corrected field equations in perturbation theory around a classical solution. The field equations are deduced via extremization of the quantum corrected action arising in effective field theory. This procedure is justified when the physical set-up of interest is semiclassical. By this we mean situations in which the wavefunctional over the space of field configurations is concentrated around the semiclassical solution, so that the path integral is well-approximated by extremization of the action. One can imagine situations where this does not hold. For example, in strongly coupled systems like QCD at nuclear energy scales, large fluctuations in the path integral lead to large corrections to this approximation. Also, it is possible that the source in the field equations is not itself semiclassical: it might be a macroscopic superposition state or otherwise lack even coarse-grained concentration in Hilbert space. However, for situations like the formation of a large black hole, or a macroscopic charge distribution, these concerns do not generally arise.


\bibitem{Calmet:2019eof}
X.~Calmet, R.~Casadio and F.~Kuipers,
Phys. Rev. D \textbf{100}, no.8, 086010 (2019)
[arXiv:1909.13277 [hep-th]].






\bibitem{Calmet:2018rkj}
X.~Calmet, B.~K.~El-Menoufi, B.~Latosh and S.~Mohapatra,
Eur. Phys. J. C \textbf{78}, no.9, 780 (2018)
[arXiv:1809.07606 [hep-th]].


\bibitem{Codello:2015pga}
A.~Codello and R.~K.~Jain,
Class. Quant. Grav. \textbf{34}, no.3, 035015 (2017)
[arXiv:1507.07829 [astro-ph.CO]].

\bibitem{Donoghue:2014yha}
J.~F.~Donoghue and B.~K.~El-Menoufi,
Phys. Rev. D \textbf{89}, no.10, 104062 (2014)
[arXiv:1402.3252 [gr-qc]].

\bibitem{Satz:2004hf}
A.~Satz, F.~D.~Mazzitelli and E.~Alvarez,
Phys. Rev. D \textbf{71} (2005), 064001
[arXiv:gr-qc/0411046 [gr-qc]].

\bibitem{Calmet:2021lny}
X.~Calmet and F.~Kuipers,
Phys. Rev. D \textbf{104} (2021) no.6, 066012
[arXiv:2108.06824 [hep-th]].


\bibitem{Donoghue:2015nba}
J.~F.~Donoghue and B.~K.~El-Menoufi,
JHEP \textbf{10} (2015), 044
[arXiv:1507.06321 [hep-th]].


\bibitem{Donoghue:2015xla}
J.~F.~Donoghue and B.~K.~El-Menoufi,
JHEP \textbf{05} (2015), 118
[arXiv:1503.06099 [hep-th]].




\end{thebibliography}
\end{document}